# Chirp spectroscopy applied to the characterization of Ferromagnetic Resonance in Magnetic Tunnel Junctions

M. Ricci[1], P. Burrascano[1], M. Carpentieri[2], R. Tomasello[3], G. Finocchio[4]
{marco.ricci,pietro.burrascano}@unipg.it, mario.carpentieri@poliba.it, tomasello@deis.unical.it,gfinocchio@unime.it

[1]Department of Engineering, Polo Scientifico Didattico di Terni, University of Perugia, Terni, TR, I-50100 Italy
[2]Department of Electrical and Information Engineering, Politecnico of Bari, via E. Orabona 4, I-70125 Bari, Italy
[3]Department of Computer Science, Modelling, Electronics and System Science, University of Calabria, Via P. Bucci, I-87036, Rende (CS), Italy
[4]Department of Electronic Engineering, Industrial Chemistry and Engineering, University of Messina, C.da di Dio, I-98166, Messina, Italy

Magnetic Tunnel Junction devices find use in several applications based on the exploitation of the Spin-Transfer Torque phenomenon. The Ferromagnetic Resonance curve is a key characteristic of any Magnetic Tunnel Junctions. It is usually characterized both experimentally and numerically by performing a lot of measurements of the magnetic response to a sinusoidal field or current. Here we propose the use of a chirp signal as excitation signal to reconstruct the Ferromagnetic resonance curve with a single measurement/simulation. A micromagnetic comparison of the proposed method with the traditional one is shown.

*Index Terms*—Ferromagnetic Resonance, Chirp signal, Magnetic Tunnel Junction, Spin Hall Effect.

## I. INTRODUCTION

Magnetic Tunnel Junction (MTJ) represents one of the most interesting structures in the field of spintronics and quantum electronics. It finds many applications such as non-volatile memory and hard drive read heads for high-density information storage. On the other hand, the discovery of the Spin-Transfer–Torque phenomena [1], [2] opened a plenty of new scenarios for the use of MTJs. In particular MgO-based MTJs, beside their application as STT-magnetic memories, are at the basis of extremely promising devices for microwave generation and detection: (a) the Spin-Transfer-Nano-Oscillator [3], [4] and (b) the Spin-Torque-Diode (STD) [5], [6],[7],[8],[9].

With respect to the latter application, recent experiments have demonstrated that nanoscale MTJs can rectify microwave signals and can work as microwave signal detectors. The rectified voltage is produced by the non-linear coupling between the magnetoresistance oscillating at microwave frequency due to the precession of the MTJ magnetization and the applied microwave signal. The maximum rectified voltage is observed when the frequency of the microwave signal is equal to the ferromagnetic resonance (FMR) frequency of the MTJ magnetization. This is one reason to characterize the FMR of MTJ devices, in order to individuate the frequency range of maximum sensibility.

The FMR is computed as magnetic response to a radio frequency (RF) magnetic field or spin transfer torque. In the last scenario, an oscillating current is applied to resonantly excite the magnetic precession. In MTJs, the oscillation of the magnetization gives rise to a tunneling magnetoresistive signal which oscillates at the same frequency. By sweeping the frequency of the microwave current, the maximum output signal of the rectified voltage produced is expected when the frequency is swept through the resonance frequency, while a very small signal is expected at frequencies far from the resonance frequency [5], [10]. A similar procedure is reproduced in numerical experiments by means of micromagnetic modeling: a sinusoidal excitation is applied to the structure for a certain long time and the steady-state amplitude of the magnetization oscillation is considered. By repeating the simulations several times at different frequencies the curve of the FMR can be reconstructed. Here, we study the possibility to use a broad band excitation, to characterize the FMR in a single measurement. Precisely, we test the method by executing micromagnetic analysis of two main structures: a standard MTJ structure, described in [11] and [12], and a 3-terminal MTJ structure deposited on a Tantalum strip that exhibits giant Spin Hall Effect (SHE), see [13],[14],[15].

Among the possible wideband excitation waveforms, the use of chirp signals has been preferred for two main reasons: (a) it allows all the frequency range of interest to be spanned continuously with a single shot excitation; (b) it allows the behavior of the MTJ as STD to be studied in presence of a complex excitation widely used in microwave radar and communication for its peculiar characteristics.

As described in the following Section, the chirp signal is characterized by an instantaneous frequency that can be defined to follow any continuous path in the time-frequency plane. By limiting ourselves to linearly span the range of interest, it is possible to associate at each excitation time a single frequency value and therefore a frequency analysis, i.e. the FMR analysis, can be executed by a time domain processing. This method is called chirp spectroscopy and it could bring to a drastic reduction of the measurement time needed to characterize the FMR, both from an experimental and a numerical point of view.

The MTJ-based STDs have been recently used for on-chip microwave interferometers to measure not only the magnitude of a microwave field but also the relative phase between the electric field **E** and the magnetic field **B** [16]. The interferometry device developed opens the realization of several microwaves imaging procedures by using both sinusoidal or multi-frequency signals. Chirp excitation could be extremely useful in this perspective for its well-known spectral and auto-correlation properties that made chirp one of the most used waveform for pulse-compression applications [17]. Moreover the Frequency Modulated Continuous Wave protocol [18] developed for chirp signal well fit with the spintronic interferometry proposed in [16] and [19]. The characterization of the magnetization dynamics of an MTJ device excited by a chirp signal, either current or field, it is therefore of interest.

At the same time, the capability to identify the FMR in several structures represents a fundamental tool to study the physics of many physical processes, for instance the determination of the SHE angle [17], [18], the dynamics of spin waves, etc.

## II. CHIRP SIGNAL

A generic chirp signal, defined in the time interval $\in [0,T]$, is represented by the expression:

$$s(t) = A\sin[\Phi(t)] = A\sin\left[2\pi\left(f_0 t + \frac{\Delta f}{2}\int_0^t x(\tau)d\tau\right)\right] \quad (1)$$

where: $A$ is the signal amplitude, $f_0$ is the central frequency, $\Delta f$ is the signal bandwidth, $x(\tau)$ is a smooth monotonic signal taking values $x(0) = -1$ and $x(T) = +1$, $\Phi(t)$ is the resultant phase function. If $\Phi(t)$ is a linear function, we retrieve the standard definition of a sinusoidal signal, while when $\Phi(t)$ is non-linear, the resulting signal is harmonic, but with a not unique oscillation frequency. In particular, unless a constant phase term, any chirp signal is completely characterized by its instantaneous frequency trajectory $f_{ist}(t)$, which is related to the phase function $\Phi(t)$ by the equation:

$$f_{ist}(t) = \frac{1}{2\pi}\frac{d\Phi(t)}{dt} = f_0 + \frac{\Delta f}{2}x(t) \quad (2)$$

The concept of the instantaneous frequency is of utmost importance in chirp design since from the trajectory of $f_{ist}(t)$ also the spectral content of the signal can be derived [22].

**FIG. 1 HERE**

With respect to the measure of the FMR, we exploited linear chirp signals that exhibit an instantaneous frequency varying with the rule: $f_{ist}(t) = f_1 + \frac{(f_2 - f_1)}{T}t$ where $f_1$ is the starting frequency, $f_2$ the stop frequency, $f_0 = \frac{f_1+f_2}{2}$ and $\Delta f = f_2 - f_1$. If $f_2 > f_1$ an "UP" chirp is obtained, conversely for $f_2 < f_1$ the chirp is named "DOWN". By the definition of the instantaneous frequency [19], the following expression can be derived for a linear chirp:

$$s(t) = A\sin\left[2\pi\left(f_1 t + \frac{\Delta f}{2T}t^2\right)\right] \quad (3)$$

from which it is straightforward to see that in (1) the linear chirp case is retrieved for $x(t) = \frac{2t}{T} - 1$.

Of course, although the instantaneous frequency is clearly defined, due to the finite length of the signal a perfect confinement of the power spectrum in the range $[f_1, f_2]$ is unrealizable. Nevertheless, if the product $T \times \Delta f$ of the duration of the signal $T$ and the bandwidth $\Delta f$ is large enough, i.e. $T \times \Delta f \gg 1$, the power spectral density is almost flat and well confined in the region $[f_1, f_2]$. An example of a linear chirp of duration $T=10$ns defined in the range 2-10 GHz is reported in Fig. 1 together with the plot of the instantaneous frequency and of the spectral amplitude. Moreover, to illustrate how the higher the $T \times \Delta f$ product is, the higher is the confinement, the spectral amplitudes of two chirps having the same $\Delta f$, but longer $T$ are shown. Due to the aforementioned properties, linear chirp signals have been used for several years to perform spectroscopy analysis: if a linear chirp is used as input of a linear system, the output signal envelope represents the magnitude of the transfer function in the spanned range. In the present case, the FMR is an inherently non-linear phenomenon but it is worth to study if the chirp excitation can represent a valid alternative to the standard procedure.

To accomplish this aim we executed various full micromagnetic simulations by exciting the structures below described with a RF current following a chirp waveform. In particular, we performed simulations by varying the duration, the bandwidth and the amplitude of chirp current waveforms. Furthermore, we compare the results obtained by using UP and DOWN chirps. This comparison allows highlighting the non-linear feature of the process, indeed, if the system was linear, the same spectral responses should be measured for both UP and DOWN chirps.

## III. NUMERICAL MODEL

We compute the FMR by using the presented method on two different strctures:
(A): an MTJ composed by a synthetic antiferromagnet (SAF) pinned layer (PL) [IrMn(6.1)/CoFe(1.8)/Ru/CoFeB(2.0)], tunnel barrier [MgO(1.25)], magnetic free layer (FL) [CoFe(0.5)/CoFeB(3.4)] (the dimensions are in nm) with elliptical cross section of 65x130 nm$^2$ (see Fig. 2a).

**FIG. 2 HERE**

(B): a SHE-MTJ composed by an MTJ device CoFeB(1)/MgO(1.2)/CoFeB(4)/Ta(5)/Ru(5) (thicknesses in nm) deposited on a Tantalum strip (6000 x 1200 x 6 nm$^3$). The thinner CoFeB (1) is the free layer and it is coupled to the Ta strip, while the thicker CoFeB (4) acts as a pinned layer, with a magnetization oriented along the $-y$-direction (see Fig. 2b). Because of the ultra-thin free layer, we take into account a very high perpendicular anisotropy ($K_u$=0.9x10$^6$ J/m$^3$) and the interfacial Dzyaloshinskii-Moriya Interaction (DMI), due the coupling between the Tantalum strip (heavy metal with a large spin-orbit coupling) and the ferromagnetic free layer [23], [24]. The initial state of the free layer magnetization is out-of-plane in the positive $z$-direction. The structure allows us to inject an in-plane current $J_{Ta}$ (which gives rise to the spin-Hall effect) through the Tantalum strip and a perpendicular current $J_{MTJ}$ via the MTJ stack. We perform micromagnetic simulations based on the numerical solution of the Landau-Lifshitz-Gilbert-Slonczewski (LLGS) equation [2], where the standard effective field, the magnetostatic field due to the polarizer, and the Oersted field due to the current are taken into account. In the case of the structure (A) the equation is described in [25] typical parameters for the free-layer were: saturation magnetization $M_S$=1000x10$^3$ A/m, damping constant $\alpha$=0.01, exchange constant $A$=2.0×10$^{-11}$ J/m, uniaxial anisotropy 4x10$^3$ J/m$^3$, spin polarization 0.6. The free layer has been discretized in computational cells of 5x5x4 nm$^3$. Regarding the structure (B), the spin-orbit torque term driven by SHE is also included in the LLGS [15]. Magnetic parameters are: $M_S$=1000x10$^3$A/m, $A$=2.0x10$^{-11}$J/m, $\alpha$=0.015, spin angle $\alpha_H$ =-0.15, and spin-polarization $\eta_T$ =0.66. The description of the model lies beyond the scope of the present paper and the reader can found exhaustive details on the literature, see for example [25] , [26] . In both cases, the integration time step is 32 fs. The simulations have been executed by using the time step setting the temperature at $T$=1K and the chirps used spanned the range 4-8 GHz and 13-20 GHz for the structures (A) and (B) respectively.

## IV. RESULTS AND DISCUSSION

Scenario (A). Several chirp current waveform were simulated. Fig. 3 shows the comparison between the FMR curve obtained by using sine excitation ($FMR_{Sine}$) and the FMR curves attained by using UP and DOWN chirps ($FMR_{UP}$ and $FMR_{DOWN}$). At a first sight, it can be noted that $FMR_{Sine}$ and $FMR_{UP/DOWN}$ ones are similar even if, as expected, the magnetization dynamics shows a dependence on the excitation frequency trajectory due to the non-linear nature of the process, i.e. the direction of the chirp is important. Indeed the curves for UP and DOWN chirp reach the peak for different f values, ($f^*_{UP}$ and $f^*_{DOWN}$). In particular we found that $f^*_{UP}$ and $f^*_{DOWN}$ are almost symmetrical displaced with respect to the value $f^*_{SINE}$ calculated by the standard method but in general Chirp UP, with instantaneous frequency increasing with time, achieves results closer to the traditional FMR curve than those provided by Chirp DOWN. We note also that the FMR linewidth Γ, i.e. the width of the resonance at half of the maximum, is larger for chirp UP and smaller for chirp DOWN, i.e. $\Gamma_{UP} > \Gamma_{DOWN}$. These facts suggest that the non-linear nature of the process hampers the rise of the magnetization oscillations so that it is necessary a certain excitation time to reach their maximum. The result is a delay in the Chirp response and therefore a blu-shift of the peak frequency in the UP case and a red-shift in the DOWN case. Moreover this phenomenon is stronger (the delay is longer) for the DOWN chirp, so that the relative linewidth is narrower.

**FIG. 3 HERE**

The results obtained by varying the chirp duration and the excitation current amplitude are summarized in Fig. 4-a where the trends of $f^*_{SINE}$, $f^*_{UP}$ and $f^*_{DOWN}$ versus the current amplitude and the duration T are reported. It can be also noted that the longer the chirp duration is, the closer are the $FMR_{UP}$ and $FMR_{DOWN}$ values to $FMR_{Sine}$ and a similar trend is attained by increasing the excitation current amplitude. At the same time, as the current increases, the resonance frequency decreases. To better evaluate how close the $FMR_{UP/DOWN}$ curves are with respect the $FMR_{Sine}$, we introduce an "accuracy" parameter defined as the ratio between the displacement between the $FMR_{UP(DOWN)}$ and $FMR_{Sine}$:

$$Accuracy = \frac{FMR_{UP/DOWN} - FMR_{Sine}}{\Gamma_{Sine}} \quad (4)$$

**FIG. 4 HERE**

The accuracy calculated for Chirp-UP and -DOWN at different time duration and current are reported in Fig. 4-b,c, from which we can conclude that, for the estimation of the FMR for the MTJ structure, Chirp-UP achieves in general better results than Chirp-DOWN and it can represent a valid alternative to the traditional sine excitation. It allows the reduction of computational time as well it provides a higher frequency resolution in the peak estimation. Furthermore, if a better accuracy is required, one can estimate the FMR peak frequency by taking the average of the results obtained with UP and DOWN chirps: $FMR_{Chirp}$=( $FMR_{UP}$ + $FMR_{DOWN}$)/2 . Fig. 5 reports the values of $FMR_{Chirp}$ derived by the data shown in Fig. 4 and the relative accuracy. It is found that $FMR_{Chirp}$ exhibits an accuracy lower than 1 in all cases even for T=10ns. It is therefore possible to estimate the FMR peak frequency with good accuracy with two simulations of 10 ns employing a chirp UP and a chirp DOWN respectively.

**FIG. 5 HERE**

To further confirm this result, we performed similar analysis on the SHE-MTJ structure aforementioned. The curves of the FMR reconstructed by using sine and chirp excitation are reported in Fig. 6. Also in this case we found that the FMR curves obtained with chirp excitations are symmetrically shifted with respect to the resonance frequency individuated by means of sine excitation. As for the MTJ case, the UP chirp –based FMR curve exhibits a shift toward higher frequencies while DOWN chirp-based FMR curve is shifted toward lower frequency and, also in this case, the UP chirp FMR curve is very close to the sine-based FMR curve. Moreover, for the UP case, sharp peaks are found near the FMR, that are not detected by the sine-based method. The presence of these peaks will be better investigated in order to link to peculiar physical phenomena. Fig 6 shows also the comparison of the FMR curve obtained by considering or not the DMI interaction: quite generally the FMR curve depends on this term but, for structures with circular shape as in this case, the FMR is not influenced by this term.

**FIG. 6 HERE**

## V. CONCLUSION

A numerical procedure for the characterization of the FMR response of MTJs exploiting chirp coded excitation has been presented. The results show that the chirp-based approach can be a valid and effective alternative of the traditional procedure allowing a significant reduction of the computational time. Moreover, the use of chirp signal has highlighted the peculiar non-linear behavior of the FMR phenomenon and it is of interest to gather further information about this process. In this perspective, it has been proposed in literature to use a non-linear exponential chirp as a input signal to characterize non-linear systems [28] and this method could be applied also for the characterization of the FMR. The use of chirp excitation is also interesting from an experimental point of view since it can find application to characterize and exploit spin-diode and spin-oscillator MTJ-based structures.

### ACKNOWLEDGMENT

This work was supported by the project PRIN2010ECA8P3 from Italian MIUR. University of Perugia and University of Calabria acknowledge financial support from Fondazione

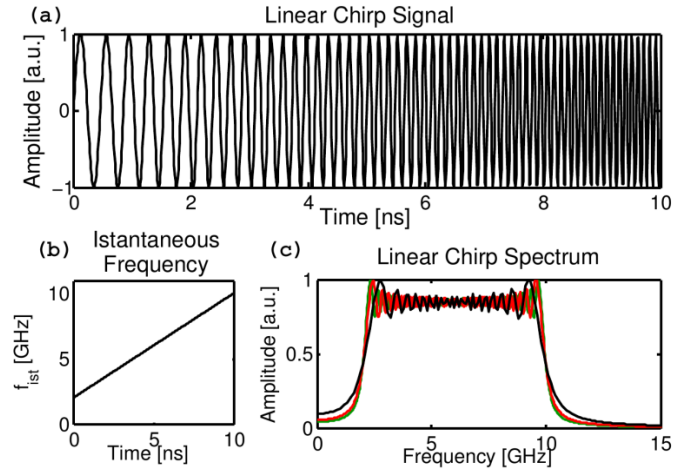

Fig. 1. (a) Example of linear chirp signal; (b) its instantaneous frequency, (c) spectral amplitude of a linear chirp for $T$=10, 30 and 50 ns, black, red and green curves respectively.

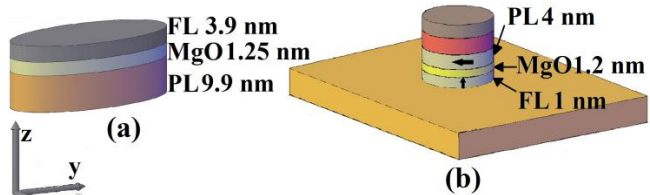

Fig. 2. Sketch of the structures under investigation (a) MTJ device and (b) SHE-MTJ device.

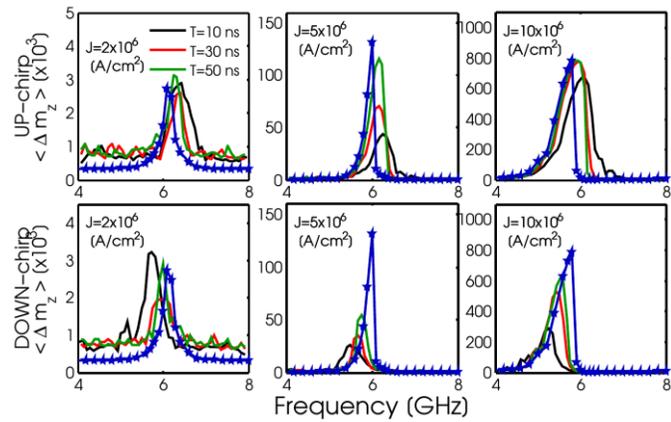

Fig. 3. Comparison between FMR calculated with sine excitation at various current amplitudes and with UP and DOWN chirp signals.

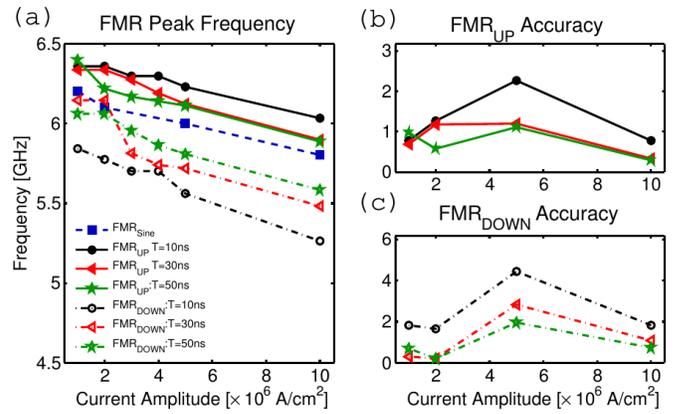

Fig. 4. $FMR_{UP}$ and $FMR_{DOWN}$ peak frequency and relative accuracy at various current amplitudes and chirp duration.

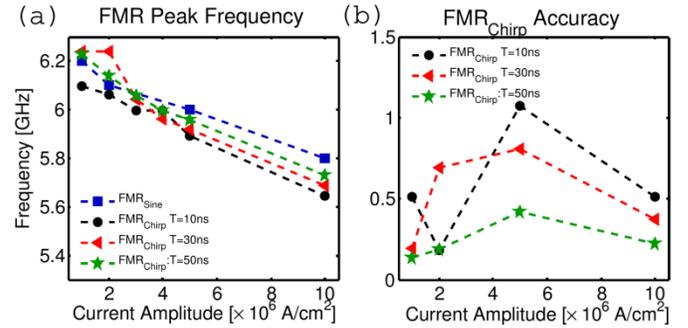

Fig. 5. $FMR_{Chirp}$ peak frequency and relative accuracy at various current amplitudes and chirp duration.

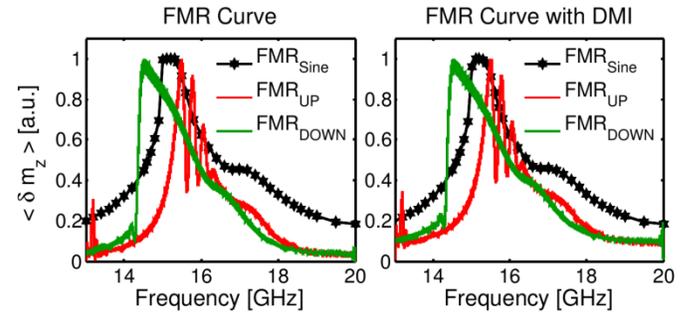

Fig. 6. Comparison between FMR curve calculated with sine excitation and chirp excitation without (left) and with (right) Dzyaloshinskii-Moriya Interaction